\documentstyle[prl,aps,multicol]{revtex}
\begin{document}
\author{K.L. Sebastian}
\address{Department of Inorganic and Physical Chemistry\\
Indian Institute of Science\\
Bangalore 560012\\
India}
\title{KINK MOTION AND THE KRAMERS PROBLEM FOR A CHAIN MOLECULE }
\maketitle
\begin{abstract}
We consider the generalization of the Kramers escape over a barrier problem
to the case of a long chain molecule. It involves the motion chain molecule
of $N$ segments across a region where the free energy per segment is higher,
so that it has to cross a barrier. We use the Rouse model and find that the
free energy of activation has a square root dependence on the temperature
 $T$, leading to a non-Arrhenius form for the rate. We also show that
there is a
special time dependent solution of the model, which corresponds to a kink in
the chain, confined to the region of the barrier. The polymer goes from one
side to the other by the motion of the kink in the reverse direction. If
there is no free energy difference between the two sides of the barrier,
then the kink moves by diffusion and the time of crossing $t_{cross}\sim
N^2/T^{3/2}$. If there is a free energy difference, then the kink moves with
a non-zero velocity from the lower free energy side to the other, leading to 
$t_{cross}\sim N/\sqrt{T}$.
\end{abstract}
The problem of thermally activated escape of a particle over a barrier (the
Kramers problem) is very well studied (see the review \cite{Hanggi}). Here
we consider a generalization of the problem to the motion of a long chain
molecule over a region where its segments have a higher free energy. The
simplest such situation is one in which a polymer goes across the interface
of two immiscible liquids \cite{CP}. Recent simulations on this problem
found that a peptide, when placed in the aqueous phase near a water-hexane
interface rapidly translocates toward the hexane phase. A more complex and
interesting situation is one in which a biomolecule moves from one side of a
biological membrane to the other \cite{HR}. The questions that we study are:
given the functional form of the barrier, (i) what is the free energy of
activation for polymer entry into the barrier? (ii) once a part of the
polymer has entered the barrier, how much time would it take for the entire
polymer to move across it? We denote this time by $t_{cross}$. Muthukumar
and Baumgartner \cite{MB} studied the effect of entropic barriers on
dynamics of polymers. They found the total passage time to depend
exponentially on $N$. Park and Sung \cite{PS} have considered the effect of
a small free energy difference between the two sides of the barrier. For a
flexible molecule they argue that $t_{cross}$ scales as $N^3$, but this can
go over to $\sim N^2$ if there is a free energy difference between the two
sides. We consider a polymer undergoing activated crossing over a barrier
whose width $w$ is larger than the Kuhn length $l$ of the polymer, but small
in comparison with the length of the polymer. Thus, we assume $l<<w<<Nl$ .
In comparison, the entropic barrier \cite{MB}\cite{PS} has a rather large
width ($\sim Nl)$. We show that for a symmetric barrier, $t_{cross}\sim
N^2/T^{3/2}$~ while if there is a free energy difference between the two
sides, then $t_{cross}\sim N/\sqrt{T}$.

We consider a one dimensional model for both barrier and the chain molecule,
described by the continuum version of the Rouse model \cite{DE}

\begin{equation}
\label{one}\zeta \partial _tR(n,t)=m\partial _{nn}R(n,t)-V^{\prime
}(R(n,t))+f(n,t) 
\end{equation}
$R$ denotes the position of the $n^{th}$ unit of the polymer at the time $t$%
. The boundary conditions for the above equation are $\left\{ \partial
_nR(n,t)\right\} _{n=0}=\left\{ \partial _nR(n,t)\right\} _{n=N}=0$. $V(R)$
is the free energy per unit of the polymer at the position $R$. The portions
of the chain over the barrier have a higher free energy per unit. The
barrier is assumed to be localized near the origin, and to extend in space
from $-a_0$ to $a_1$, with $a_0<a_1$. We take $V(R)$ to be a smooth
continuous function, with $V^{\prime }(R)=2\,k\,R\left( R+a_0\right)
\,\left( R-a_1\right) $ for $-a_0<R<a_1$and $V^{\prime }(R)=0$ outside these
limits. Integrating this and choosing the potential to be zero for $R<-a_0$,
we get 
\begin{equation}
\label{three}V(R)=\frac k6(R+a_0)^2(3R^2-2Ra_0-4Ra_1+a_0^2+2a_0a_1) 
\end{equation}
for $-a_0<R<a_1$ and $V(R)=\,\frac{\,1}6k\left( a_0-a_1\right) \,\left(
a_0+a_1\right) ^3$ for $R>a_1$. The barrier height for the forward crossing
is $V_f=\frac{\,1}6ka_0{}^3\,\left( a_0+2\,a_1\right) $ and for the reverse
process, it is $V_b=\frac{1\,}6ka_1{}^3\left( 2\,a_0+a_1\right) $. On
crossing the barrier, a unit of the polymer lowers its free energy by $%
\Delta V=\frac{1\,}6k\left( a_0-a_1\right) \,\left( a_0+a_1\right) ^3$. The
form of the potential is shown in the figure.Though our results are for this
particular form of the potential, the conclusions are general and
independent of the form of the potential.

$f(n,t)$ are random forces acting on the $n^{th}$ segment and have the
correlation function $\left\langle f(n,t)f(n_1,t_1)\right\rangle =2\zeta
k_BT\delta (n-n_1)\delta (t-t_1)$. At equilibrium, the probability
distribution functional is $\exp \left[ -\frac 1{k_BT}\int dn\left\{ \frac 12%
m\left( \partial _nR\right) ^2+V(R(n))\right\} \right] $. The free energy of
activation for the crossing can be found by extremising this functional,
subject to the condition that one end of the polymer has crossed $R_{\max }$%
, the maximum of the barrier. Assuming the polymer to be very long, we find
this by taking $R(-\infty )=-a_0$ and the other end of the polymer to be at
a point with $R>R_{\max }$. This extremum configuration satisfies the
Newton's equation $m\partial _{nn}R=V^{\prime }(R)$, subject to these
conditions. For this Newton's equation, the conserved energy is $E_c=\frac 12%
m\left( \partial _nR\right) ^2-V(R(n))$ and for the extremum path its value
is equal to zero. The particle starts at $R(-\infty )=-a_0$ and ends up at $%
R_f$ where $R_f(>R_{\max }),$ is the point such that $V(R_f)=0$. We find $%
R_f=a_0(\gamma -\,\sqrt{\gamma ^2-\gamma })$ where $\gamma =(1+2\frac{a_1}{%
a_0})\frac 13$. The free energy of activation is $E_{act}=\int_{-a_0}^{R_f}%
\sqrt{2mV(R)}dR$. Interestingly, in the Rouse model, the parameter $m$ is
proportional to the temperature \cite{DE}. So if the $V(R)$ is temperature
independent, $E_{act}$ $\alpha \sqrt{T}$. One evaluation, we find $E_{act}=%
\frac{\sqrt{mk}a_0^3}6\left[ (3\,\gamma ^2+1)\sqrt{1+3\gamma }-3\,\gamma
\,(\gamma ^2-1)\ln \left( \sqrt{\gamma (\gamma -1)}/\left( 1+\gamma -\sqrt{%
1+3\gamma }\right) \right) \right] $.

The Boltzmann factor $e^{-\frac{E_{act}}{k_BT}}$ for the crossing of one end
of the polymer over the barrier thus has the form $e^{-cons\tan t/\sqrt{T}}$
and is independent of $N$ for large $N$.

Now, to calculate $t_{cross}$, we first look at the mathematical solutions
of the deterministic equation, in which we replace random noise term $f(n,t)$
by its average. The result is

\begin{equation}
\label{four}\zeta \partial _tR(n,t)-m\partial _{nn}R(n,t)+V^{\prime }(R)=0 
\end{equation}
with $\left\{ \partial _nR(n,t)\right\} _{n=0}=\left\{ \partial
_nR(n,t)\right\} _{n=N}=0$. The simplest solutions of this equation are: $%
R(n,t)=R_0$ ( $R_0$ is a constant), with $R_0<-a_0$, or with $R_0>a_1$.
These correspond to the polymer being on either side of the barrier. Thermal
noise makes $R(n,t)$ fluctuate about the mean position $R_0$ and this may be
analyzed using the normal co-ordinates for fluctuations about this mean
position. Each normal mode (phonon) obeys a Langevin equation similar to
that for a harmonic oscillator, executing Brownian motion. Further, the mean
position (center of mass) $R_0$ itself executes Brownian motion\cite{DE}. In
addition to these two time independent solutions, the above equation has a
time dependent solution (a kink) too, which corresponds to the polymer
crossing the barrier. We analyze the dynamics of the chain, with the kink in
it, using the normal modes for fluctuations about this kink configuration.
Our analysis makes use of the techniques that have been used to study the
diffusion of solitons\cite{solitons}

The kink solution may be found using the ansatz $R(n,t)=R_s(\tau )$ where $%
\tau =n-vt$ \cite{solitons}, when the equation (\ref{four}) reduces to $m\,%
\frac{d^2R_s}{d\tau ^2}+v\,\zeta \,\frac{dR_s}{d\tau }=V^{\prime }(R_s)$. If
one imagines $\tau $ as time, then this is a simple Newtonian equation for
the motion of particle of mass $m$, moving in the upside down potential $%
-V(R)$ , and $v\,\zeta /m$ is the coefficient of friction. We can easily
find a solution of this equation, obeying the conditions $R_s(\tau )=-a_0$
for $\tau \longrightarrow -\infty $ and $R_s(\tau )=a_1$for $\tau
\longrightarrow \infty .$ It is $R_s(\tau )=\left( -a_0+e^{\tau \,\omega
\,\left( a_0+a_1\right) }\,a_1\right) \left( 1+e^{\tau \omega \,\left(
a_0+a_1\right) }\right) ^{-1}$ , with $\omega \,=\sqrt{k/m}$ and the velocity%
{\ }$v=\frac{\sqrt{mk}\,}\zeta \,(a_0-a_1).$ The solution is a kink,
occurring in the portion of the chain inside the barrier. The point with $%
\tau =0$ shall be referred to as the center of the kink. (Actually one has a
one-parameter family of solutions of the form $R_s(\tau +\tau _0)$, where $%
\tau _0$ is any arbitrary contant). As $\tau =n-vt$, the center of the kink
moves with a constant velocity $v$, which depends on the barrier heights.
Note that if $a_0<a_1$, then $V_f<V_b$, and this velocity is negative. This
implies that the kink is moving in the negative direction, which corresponds
to the chain moving in the positive direction. That is, the chain moves to
the lower free energy region, with this velocity. If the barrier is
symmetric $a_0=a_1$( $V_f=V_b$) then the velocity of the kink is zero.

The center of the kink can be anywhere on the chain and hence the kink can
move on the chain. It would also execute Brownian motion, due to thermal
fluctuations of the medium. To analyze this, following `Instanton methods'
of field theory \cite{RR}, we write 
\begin{equation}
\label{nine}R(n,t)=R_s(n-a(t))+\sum_{p=1}^\infty X_p(t)\phi _p(n-a(t),t) 
\end{equation}

We have allowed for the motion of the kink by allowing the kink center to be
at $a(t)$, where $a(t)$ is a random function of time which is to be
determined. $\phi _p$ are a set of functions defined below and $X_p(t)$ are
expansion coefficients. $\phi _p$ eventually will turn out to describe
motion along the $p^{\text{th}}$ normal mode in presence of the kink in the
chain. Using the equation (\ref{nine}) in the equation (\ref{one}) and
expanding around the kink, retaining first order terms in $X_p(t)$, and
changing over to the new variables $(\overline{n},t)$ with $\overline{n}%
=n-a(t)$, (which is just segment position along the chain with respect to
the center of the kink) we get

\begin{equation}
\label{ten}\zeta (v-\stackrel{.}{a}(t))\partial _{\overline{n}}R_s(\overline{%
n})+\zeta \sum_{p=1}^\infty \stackrel{\cdot }{X}_p(t)\phi _p(\overline{n}%
,t)+\sum_{p=1}^\infty X_p(t)\widehat{{\cal L}}\phi _p(\overline{n},t)=f(%
\overline{n}+a(t),t), 
\end{equation}
where $\widehat{{\cal L}}=\zeta \partial _t-m\partial _{\overline{n}%
\overline{n}}-\stackrel{.}{\zeta a}(t)\partial _{\overline{n}}+V^{\text{ }%
^{\prime \prime }}(R_s(\overline{n}))$. We take $\phi _p$ to obey the
equation $\left\{ \zeta \partial _t-m\partial _{\overline{nn}}-\zeta
v\partial _{\overline{n}}+V^{\text{ }^{\prime \prime }}(R_s(\overline{n}%
))\right\} \phi _p(\overline{n},t)=0$, subject to the boundary conditions $%
\partial _{_{\overline{n}}}\phi _p(\overline{n},t)=0$ at both ends of the
chain. (i.e. at $\overline{n}=-a(t)$ and at $\overline{n}=N-a(t)$). For a
very long chain, with the kink located well inside it, the ends cannot
influence the dynamics of barrier crossing. Therefore, we would make only a
minor error by imposing these boundary conditions at $\overline{n}=\pm N/2$.
Then, putting $\phi _p(\overline{n},t)=\psi _p(\overline{n})e^{-\lambda
_pt/\zeta -v\zeta \overline{n}/(2m)}$, we find that $\psi _p(\overline{n})$
has to obey $\widehat{H}\psi _p(\overline{n})=\lambda _p\psi _p(\overline{n}%
) $ with $\widehat{H}=\left\{ -m\partial _{\overline{n}\overline{n}%
}+V^{^{\prime \prime }}(R_s(\overline{n}))+(v\zeta )^2/(4m)\right\} $. The
translation mode of the kink leads to a solution $\psi _0(\overline{n})$ of
the above equation with $\lambda _0=0$ \cite{RR}. Its functional form may be
found by putting $e^{v\zeta \overline{n}/(2m)}\partial _{\overline{n}}R_s(%
\overline{n})=\omega \,\left( a_0+a_1\right) ^2e^{\frac 12\overline{n}%
\,\omega \,\left( 3\,a_0+a_1\right) }\left( 1+e^{\overline{n}\,\omega
\,\left( a_0+a_1\right) }\right) ^{-2}=C\psi _0(\overline{n})$, where is $C$
the is the normalization factor, chosen such that $\left\langle \psi _0\mid
\psi _0\right\rangle =1$. On evaluation, we find $C^2=\left\langle \partial
_{\overline{n}}R_s(\overline{n})\left| e^{v\zeta \overline{n}/m}\right|
\partial _{\overline{n}}R_s(\overline{n})\right\rangle =\frac{2\,}3\pi
\,\omega \,\csc (2\,\pi \frac{\,a_1-a_0}{a_0+a_1})\,\left( a_1-a_0\right)
\,a_0\,a_1$. The eigenfunctions $\psi _p(\overline{n}),p=0..\infty $ form an
orthonormal set.

The equation (\ref{ten}) thus becomes 
\begin{equation}
\label{thirteen}\sum_{p=1}^\infty \stackrel{\cdot }{X}_p(t)\psi _p(\overline{%
n})e^{-\lambda _pt/\zeta }+\left( v-\stackrel{\cdot }{a}(t)\right) Y(t)=%
\frac 1\zeta e^{v\zeta \overline{n}/(2m)}f(\overline{n}+a(t),t) 
\end{equation}
where $Y(t)=\psi _0(\overline{n})C+\sum_{p=1}^\infty X_p(t)e^{-\lambda
_pt/\zeta }\widetilde{\partial }_{_{\overline{n}}}\psi _p(\overline{n})$,
with $\widetilde{\partial }_{_{\overline{n}}}=e^{v\zeta \overline{n}%
/(2m)}\partial _{\overline{n}}e^{-v\zeta \overline{n}/(2m)}$. On taking the
inner product of this with $\psi _0(\overline{n})$, we get

\begin{equation}
\label{fourteen}\left( v-\stackrel{\cdot }{a}(t)\right) \left(
C+\sum_{p=1}^\infty X_p(t)e^{-\lambda _pt/\zeta }\left\langle \psi _0\mid 
\widetilde{\partial }_{_{\overline{n}}}\psi _p(\overline{n})\right\rangle
\right) =\xi _0(t) 
\end{equation}
with $\xi _0(t)=\frac 1\zeta \int_{-N/2}^{N/2}d\overline{n}\psi _0(\overline{%
n})e^{v\zeta \overline{n}/(2m)}f(\overline{n}+a(t),t)$. Obviously, $%
\left\langle \xi _0(t)\right\rangle =0$ and $\left\langle \xi _0(t)\xi
_0(t_1)\right\rangle =\delta (t-t_1)(2k_BT/\zeta )\int_{over\text{ }the\text{
}kink}d\overline{n\text{ }}e^{v\overline{n}\zeta /m}\left[ \psi _0(\overline{%
n})\right] ^2$ $=\delta (t-t_1)k_BT/(2\zeta a_0\,a_1)\sec (2\,\pi \frac{%
\,a_1-a_0}{a_0+a_1})$ $\,\,\left( 3\,a_1-a_0\right) \,\left(
3\,a_0-a_1\right) $. On taking the inner product of the equation (\ref
{thirteen}) with $\psi _l(\overline{n})$, we get

\begin{equation}
\label{seventeen}\stackrel{\cdot }{X}_le^{-\lambda _lt/\zeta }+\left( v-%
\stackrel{\cdot }{a}(t)\right) \sum_{p=1}^\infty X_p(t)e^{-\lambda _pt/\zeta
}\left\langle \psi _l\left| \widetilde{\partial }_{_{\overline{n}}}\right|
\psi _p\right\rangle =\xi _l(t) 
\end{equation}
where $\xi _p(t)=\frac 1\zeta \int_{-N/2}^{N/2}d\overline{n}\psi _l^{*}(%
\overline{n})e^{v\zeta \overline{n}/(2m)}f(\overline{n}+a(t),t)$. The
equation (\ref{seventeen}) may be solved, and using the kets $\left|
X(t)\right\rangle =\sum_{p=1}^\infty X_p(t)\left| \psi _p\right\rangle $ and 
$\left| \xi (t)\right\rangle =\sum_{p=1}^\infty \xi _p(t)\left| \psi
_p\right\rangle $, the solution is: 
\begin{equation}
\label{eighteen}\left| X(t)\right\rangle =\left| X_d(t)\right\rangle +\left|
X_r(t)\right\rangle 
\end{equation}
with $\left| X_d(t)\right\rangle =\widehat{T}\left[ \exp \left\{
\int_0^tdt_1\left( \stackrel{\cdot }{a}(t_1)-v\right) \widehat{A}%
(t_1)\right\} \right] \left| X(0)\right\rangle $, $\left|
X_r(t)\right\rangle =$ $\int_0^tdt_1\widehat{T}\left[ \exp \left\{
\int_{t_1}^tdt_2\left( \stackrel{\cdot }{a}(t_2)-v\right) \underline{%
\widehat{A}}(t_2)\right\} \right] $ $e^{\widehat{H}t_1}\left| \xi
(t_1)\right\rangle ,$where $\widehat{A}(t)$ is the operator $\widehat{A}%
(t)=e^{\widehat{H}t/\zeta }\widetilde{\partial }_{_{\overline{n}}}e^{-%
\widehat{H}t/\zeta }$. $\widehat{T}$ is the time ordering operator. The
equation (\ref{fourteen}) gives

\begin{equation}
\label{final}\stackrel{\cdot }{a}(t)=v+\xi _0(t)/\left( C+\left\langle \psi
_0\right| \widetilde{\partial }_{_{\overline{n}}}e^{-\widehat{H}t/\zeta
}\left| X(t)\right\rangle \right) 
\end{equation}
Now taking the average of the above, and using the equation (\ref{eighteen})
and the fact that $\xi _0(t)$ is independent of everything that occurred at
previous instants, we get $\left\langle \stackrel{\cdot }{a}(t)\right\rangle
=v$, so that on an average, the kink moves with a velocity $v$. For the
polymer to cross the barrier, the kink has to go in the reverse direction,
by a distance proportional to the length of the chain. Hence $t_{cross}\sim
N/v$. As $v$ is proportional $\sqrt{mk}$, assuming $V(R)$ to be temperature
independent we find $t_{cross}\sim N/\sqrt{T}$.

If the barrier is symmetric, the kink moves with an average velocity $v=0$.
If one neglects the kink-phonon scattering term in the equation (\ref{final}%
) (i.e., the term $\left\langle \psi _0\right| \widetilde{\partial }_{_{%
\overline{n}}}e^{-Ht/\zeta }\left| X(t)\right\rangle $), then the diffusion
coefficient of the kink is $D=\frac{3\,k_BT\,\,\,}{8\,\pi \,\zeta \,}\sqrt{%
\frac mk}\frac{\left( 3\,a_{1-}a_0\right) \,\left( 3\,a_0-a_1\right) }{%
\,a_0{}^2\,a_1{}^2\,\left( a_1-a_0\right) }\tan (2\pi \frac{a_1-a_0}{a_0+a_1}%
)$. In the limit $a_1\rightarrow a_0$, we get the value for the symmetric
barrier as $D=\frac{3\,k_BT\,}{4\,\zeta \,\,a_0{}^3}\sqrt{\frac mk}.$ In
this case, the time required for the polymer to cross over the barrier is $%
t_{cross}\sim N^2/D\sim N^2/T^{3/2}$.

Park and Sung \cite{PS} consider the passage of a polymer through a pore and
the barrier is entropic in origin. Consequently it is very broad, the width
being of the order of $N$. Hence they consider the movement as effectively
that of the center of mass of the polymer which diffuses with a diffusion
coefficient proportional to $1/N$. As the center of mass has to cover a
distance $N$, the time that it takes is proportional to $N^3$. If there is a
free energy difference driving the chain from one side to the other, then
the time is proportional to $N^2$. In comparison, in our process the barrier
is extrinsic in origin and its width is small in comparison with the length
of the chain. The crossing occurs by the motion of the kink, which is a
localized non-linear object in the chain whose width is of the same order as
that of the barrier. As the kink is a localized object, its diffusion
coefficient has no $N$ dependence and hence our results are different from
those of Park and Sung \cite{PS}. In the case where there is no free energy
difference, our crossing time is proportional to $N^2$(in contrast to $N^3$
of Park and Sung) , while if there is a free energy difference, our crossing
time is proportional to $N$ (in contrast to $N^2$ of Park and Sung).

It is interesting to note that perhaps, the kind of behaviour discussed here
may have already been observed in the simulations of Chipot and Pohoille\cite
{CP}. It is also of interest to note that if the barrier height for the
forward crossing decreased, approaching zero, then the velocity $v$ tends to
a finite value.

I deeply indebted to Professor S.K. Rangarajan, for the encouragement that
he has given me over the years. I thank Prof. S.Vasudevan for interesting
discussions and Prof. B. Cherayil, for his comments on the manuscript.

\bf{Figure Caption:}  The barrier and its inverted form.  The barrier heights in the forward and backward directions are shown.  The dotted line represents the path that determines the activation energy

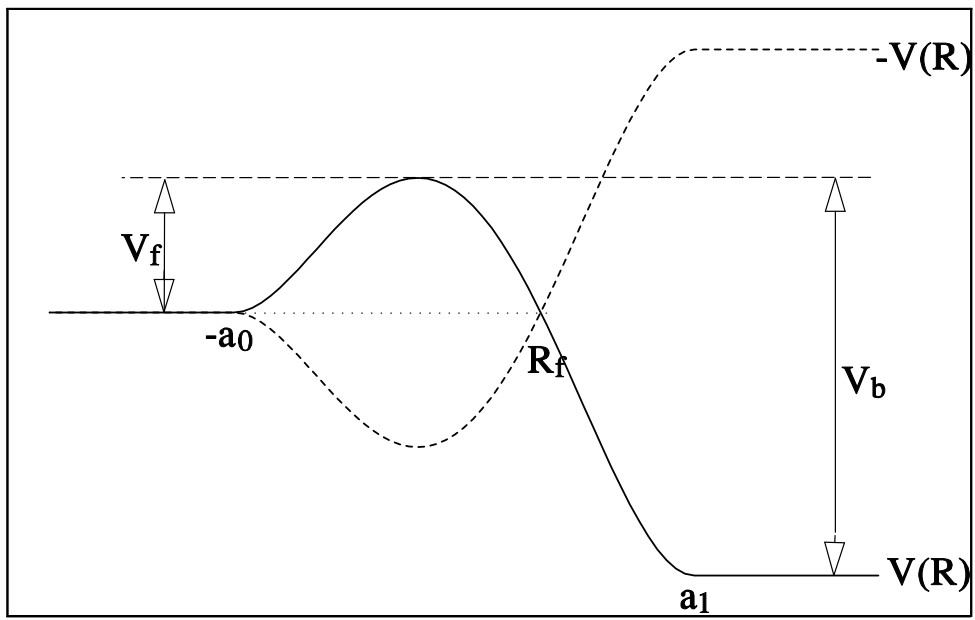
\end{document}